\documentclass[journal=jctcce,manuscript=article]{achemso}
\usepackage[utf8]{inputenc}
\usepackage[T1]{fontenc}
\usepackage[usenames,dvipsnames]{xcolor}
\usepackage[english]{babel}
\usepackage{graphicx}
\usepackage{rotating} 
\usepackage{float}
\usepackage{placeins}
\setkeys{acs}{doi = true}
\usepackage[colorlinks, linkcolor = blue, citecolor = blue, filecolor = blue,
urlcolor =blue]{hyperref}
\newcommand{\doi}[1]{\href{http://dx.doi.org/#1}{\nolinkurl{#1}}}
\usepackage{subfigure}
\usepackage{braket}
\usepackage{amsmath,mathtools}

\title[Data Hierarchies in MFML]{Investigating Data Hierarchies in Multifidelity Machine Learning for Excitation Energies}

\author{Vivin Vinod}
\affiliation{School of Mathematics and Natural Science, University of Wuppertal, 42119 Wuppertal, Germany}
\email{vinod@uni-wuppertal.de}

\author{Peter Zaspel}
\affiliation{School of Mathematics and Natural Science, University of Wuppertal, 42119 Wuppertal, Germany}

\begin{document}

\begin{abstract}
Recent progress in machine learning (ML) has made high-accuracy quantum chemistry (QC) calculations more accessible. Of particular interest are multifidelity machine learning (MFML) methods where training data from differing accuracies or fidelities are used. These methods usually employ a fixed scaling factor, $\gamma$, to relate the number of training samples across different fidelities, which reflects the cost and assumed sparsity of the data. This study investigates the impact of modifying $\gamma$ on model efficiency and accuracy for the prediction of vertical excitation energies using the QeMFi benchmark dataset. Further, this work introduces QC compute time informed scaling factors, denoted as $\theta$, that vary based on QC compute times at different fidelities. A novel error metric, error contours of MFML, is proposed to provide a comprehensive view of model error contributions from each fidelity. The results indicate that high model accuracy can be achieved with just 2 training samples at the target fidelity when a larger number of samples from lower fidelities are used. This is further illustrated through a novel concept, the $\Gamma$-curve, which compares model error against the time-cost of generating training samples, demonstrating that multifidelity models can achieve high accuracy while minimizing training data costs.
\end{abstract}

\section{Introduction}
Machine learning (ML) and quantum chemistry (QC) have become increasingly interlinked over the recent times. Both have seen rapid development in tandem allowing for quick prediction of QC properties in place of the costly conventional calculations \cite{behler2015constructing, butler_davies_review_2018, Westermayr2020review, noe2020machine, Sergei21_Chem_review_NNML, samek2021explaining}. This has allowed researchers to perform preliminary examination of complex QC problems with much speed. 
The ML-QC pipeline first identifies a QC property of interest. Next, a training set is calculated for a desired QC method, say Density Functional Theory (DFT); this is also referred to as a \textit{fidelity}, that is, the level of accuracy of the method with respect to what would be considered ground truth.
Once a training dataset is computed, a ML model of choice is trained. 

The bottleneck in such a pipeline is often the high cost of generating training data. A ML model can only be as good as the data it is trained on. 
It is often noted that a larger number of training samples results in a more accurate ML model \cite{Westermayr2020review, dral21a}. This observation implies that either one needs to use a less accurate, and thereby less expensive QC method to train the ML model, or have less training samples at a higher fidelity thereby, resulting in a less accurate ML model, when it comes to the prediction error relative to the data. 
Several methodological improvements over the single fidelity ML methods for QC have been proposed to overcome this hurdle in the ML-QC pipeline, including $\Delta$-ML \cite{Ramakrishnan2015}, where one trains on the difference between two fidelities. 
This method has been shown to reduce the number of training samples needed at the expensive fidelity and has since been modified in various flavors including hierarchical ML \cite{dral2020hierarchical}, multifidelity ML (MFML) \cite{zasp19a, patra2020multi, vinod23_MFML}, and optimized MFML (o-MFML) \cite{vinod_2024_oMFML}. MFML and its variant of o-MFML, systematically combine several $\Delta$-ML like models with more than two fidelities. This method has been shown to be superior in predicting excitation energies along molecular trajectories \cite{vinod23_MFML}. A recent work has also introduced the use of multitask Gaussian processes to harness heterogeneous multifidelity data in order to predict three-body interaction energy in water trimer with coupled cluster (CC) accuracy \cite{fisher2024multitask}.
MFML differs from the conventional $\Delta$-ML method not just in terms of the number of fidelities that are used but also in the number of training samples used at each fidelity. Conventionally, the $\Delta$-ML method uses the same number of samples at both the fidelities. In MFML, these training samples are scaled down as one increases the fidelity of the training data.
This further reduced the number of costly training samples needed at the highest fidelity, also called the \textit{target fidelity}. This \textit{scaling factor}, in previous studies was set to be 2, meaning that at each subsequently lower fidelity, the number of training samples would be scaled up by a factor of 2 \cite{zasp19a, vinod23_MFML, vinod_2024_oMFML, vinod2024_nonnestedMFML}. Ref.~\citenum{zasp19a} discusses that the scaling factor of 2 for MFML was decided based on previous work related to sparse grid combination techniques (SGCT) \cite{hegland2016combination, harbrecht2013combination, reisinger2013combination}. 

The \textit{scaling factor}, the ratio of training samples used at two consecutive fidelities, or levels, directly controls the total number of training samples used for MFML and thereby the cost of generating a training set for the approach. Understanding the effect of this parameter in the efficiency and accuracy of the MFML approach would potentially provide opportunities to further improve the overall multifidelity approach for QC.
Previously, ref.~\citenum{patra2020multi} has studied a two-fidelity MFML model with varying the number of training samples at the cheaper fidelity. By increasing the number of training samples at the lowest fidelity in an additive manner, the model error has been shown to decrease for the prediction of polymer bandgaps. 
A similar study has been performed in ref.~\citenum{Chen_2021_MFdata_bandgap} for the study of bandgaps in solids. However, these studies lack any systematic assessment of the scaling factor itself but rather loosely study the effect of training data size within a two-fidelity data structure.
This work assesses scaling factors that are different from those used thus far in literature. Several fixed scaling factors, that is, the same scaling factors across the different fidelities are systematically tested. These are evaluated on the recently released benchmarking multifidelity dataset, QeMFi \cite{vinod_2024_QeMFi_zenodo_dataset, vinod2024QeMFi_paper}, which consists of 135,000 geometries of nine complex molecules. 
Since the QeMFi dataset also provides the compute time for each fidelity for each molecule type, two time-cost informed scaling factors are also assessed. 

Studying model accuracy in relation to the cost of generating the training set for the model also provides a robust measure of how the diverse MFML models behave with respect to the single fidelity models as has been shown in refs.~\citenum{vinod23_MFML, vinod2024QeMFi_paper}. 
Therefore, assessment of model accuracy and time-cost of generating corresponding training data is made for the diverse scaling factors.
In interest of a complete investigation not only into the scaling factors but also into better understanding the multifidelity data structure, this work further introduces a new error metric for multifidelity methods for QC, namely \textit{error contours} of MFML. Error contours describe the model error with respect to training samples used at two fidelities thereby giving a more comprehensive analysis of the contribution of each fidelity to the overall accuracy of the MFML model.  
The study of the error contours of MFML indicates that using much lower training samples at the costlier fidelity while increasing the number of training samples at the lowest fidelity results in an MFML model of high accuracy at a much lower cost than the conventional MFML approach with a fixed scaling factor.
To systematically assess this, this work studies multifidelity models built with a small number of fixed training samples at the target fidelity and increasing the scaling factor. This gives rise to the notion of the $\Gamma$-curve as delineated in  \nameref{gammacurve_theory}. The models that are build in such a manner are shown to be superior to the conventional MFML approach in terms of model error for a given cost of generating the training data.

The rest of this manuscript is structured as follows: all the methodological pre-requisites including dataset details are presented in  \nameref{Methods}. The concepts of scaling factors and the tools used in this work are also explained in detail. 
This is followed by the results of MFML and o-MFML models for the prediction of excitation energies with the different scaling factors in  \nameref{results}. In addition to the time-cost of the different MFML models in  \nameref{time_cost_results}, the error contours of MFML and the $\Gamma$-curves are studied in \nameref{results_error_contours} and \nameref{gammacurves_results} respectively. 
Inferences on these results are made followed by a discussion and outlook of this work along with its implications for future work in multifidelity methods.

\section{Methods}\label{Methods}
This section discussed the various methodological pre-requisites needed to appreciate the results and inferences of this work. First, the dataset used is introduced and the multifidelity structure explained. This is followed by the ML details including KRR, MFML, and o-MFML. The section also discusses the concept of scaling factors in detail, to establish the conceptual motivation behind this work.

\subsection{Dataset}
The QeMFi dataset \cite{vinod2024QeMFi_paper, vinod_2024_QeMFi_zenodo_dataset} is a recently released benchmark dataset of diverse molecules of varying chemical conformations. These molecules include urea, 2-nitrophenol, and thymine among others. It contains a total of 135,000 point geometries with diverse QC properties such as the first 10 vertical excitation energies in addition to the compute time for each molecule for each fidelity. These properties are calculated with the DFT formalism with five different basis sets, which in turn form the ordered multifidelity hierarchy. In increasing order of accuracy, these are STO3G, 321G, 631G, def2-SVP, and def2-TZVP. In this work, these fidelities are hereon referred to only with the basis set names, and often with the shortened notation such as TZVP for def2-TZVP.

The first excitation energies were taken from QeMFi as the property to be studied with the different data scaling applied. While QeMFi provides various other QC properties such as ground state energies and molecular dipole moments, the excitation energies are chosen since they are generally more challenging for ML model \cite{Westermayr2020review, dral2020quantum} than ground state energies. Although there is increasing interest in vector properties such as molecular dipole moments, these are not studied here since they would require an extensive discussion of equivariance and invariance of molecular descriptors \cite{westermayr_2021_perspective, Veit_moldipoles_data_2020}, that is the map between the Cartesian coordinates and machine learnable input features. This discussion lies out of the scope of this current work. Thus the excitation energies of QeMFi are used herein. The energies are given in $\rm cm^{-1}$. As a first step, from the 135,000 total geometries in QeMFi, 120,000 were randomly sampled to form the diverse training sets. From the remaining 15,000 geometries, a validation set of 2,000 samples was set aside to be used in the o-MFML method. 
Finally, the remaining 13,000 samples were set aside as a test set. Both the validation set and the test set are not changed over the course of this entire study, that is, all errors are reported on the same test set.
It is to be noted that the test set is never used during the training or validation phases of any of the models. Therefore the error reported by the models is on a set of truly unseen data.
The validation set and the test set are fixed and not changed during the course of the experiments in this work.

In order to assess the efficiency of the multifidelity models vis-\'a-vis the single fidelity model, the time-cost of generating the training data versus model error are studied. For this purpose, the QC-compute time of each fidelity is considered from the QeMFi dataset. As ref.~\citenum{vinod2024QeMFi_paper} notes, the QC-compute time provided in the dataset is the time of a single-core QC-calculation of a fidelity for a given molecule. To compute the training dataset generation time, these calculation times are used. 

\subsection{Kernel Ridge Regression}
Given a fidelity, $f$, consider a training set of molecular representations (denoted as $\boldsymbol{X}_q)$ and corresponding excitation energies (denoted as $y_q^{(f)}$) given by $\mathcal{T}^{(f)}:=\{(\boldsymbol{X}_i,y^{(f)}_i)\}_{i=1}^{{N}_{\rm train}}$.
The molecular representations used in this assessment are the unsorted Coulomb Matrix(CM) \cite{Rup12CM} representations. For a given molecule, these are calculated as
\begin{equation}
C_{i,j}:=
    \begin{cases}
        \frac{Z_i^{2.4}}{2}~,&i=j\\
        \frac{Z_i\cdot Z_j}{\left\lVert \boldsymbol{R}_i-\boldsymbol{R}_j\right\rVert}~,&i\neq j~,
    \end{cases}
    \label{CM_eq}
\end{equation}
where, $Z_i$ is the atomic charge for the $i^{\rm th}$ atom and its Cartesian coordinates are given by $\boldsymbol{R}_i$. The resulting representation for a molecule is then flattened into a 1-D array.
Since the CM representation is symmetric for the atomic indices, a molecule with $m$ atoms is therefore represented by a 1-D array of size $m(m+1)/2$.
The QeMFi dataset consists of molecules of differing number of atoms, therefore the CM are padded with zero to match with the size of the CM representation of the largest molecule of the dataset which is o-HBDI with 23 atoms. This corresponds to a padded representation size of 253 entries per geometry.

For a query CM representation given as $\boldsymbol{X}_q$, the prediction of energies at fidelity $f$ are given by:  
\begin{equation}
    P^{(f)}_{\rm KRR}\left(\boldsymbol{X}_q\right) := \sum_{i=1}^{N^{(f)}_{\rm train}} \alpha^{(f)}_i k\left(\boldsymbol{X}_q,\boldsymbol{X}_i\right)~,
    \label{eq_KRR_def}
\end{equation}
where, $k(\cdot,\cdot)$ is the kernel function. In this work, the Mat\'ern kernel of first order and discrete $L_2$ norm is used across the KRR models built. The Mat\'ern kernel for two CM representations is computed as 
\begin{equation}
    k\left(\boldsymbol{X}_i,\boldsymbol{X}_j\right) = \exp{\left(-\frac{\sqrt{3}}{\sigma}\left\lVert \boldsymbol{X}_i-\boldsymbol{X}_j\right\rVert_2^2\right)}\cdot\left(1+\frac{\sqrt{3}}{\sigma}\left\lVert \boldsymbol{X}_i-\boldsymbol{X}_j\right\rVert_2^2\right)~,
    \label{eq_matern}
\end{equation}
where, $\sigma$ is the kernel length-scale parameter.
The coefficients of KRR, $\boldsymbol{\alpha}^{(f)}$ from Eq.~\eqref{eq_KRR_def}, are computed by solving 
\begin{equation}
(\boldsymbol{K}+\lambda \boldsymbol{I}) \boldsymbol{\alpha}^{(f)} = \boldsymbol{y}^{(f)}
\label{eq_solving_KRR}
\end{equation}
with $\boldsymbol{K}$ called the kernel matrix whose entries are simply the result of the kernel function from Eq.~\eqref{eq_matern} for each molecule pair in the training set. In Eq.~\eqref{eq_solving_KRR}, $\lambda$ is a parameter that penalizes overfitting.

\subsection{Multifidelity Machine Learning} \label{MFML methods}
Refs.~\citenum{zasp19a, vinod_2024_oMFML} have shown that the multifidelity machine learning (MFML) model for some QC property can be built by systematically combining  individual \textit{sub-models} that are themselves trained for a given fidelity, $f$.
Within this approach, each sub-model is then identified by the fidelity and the corresponding number of training samples used, $N_{\rm train}^{(f)}$. This is carried out by a composite index, $\boldsymbol{s}=(f,\eta_f)$, where $2^{\eta_f}=N_{\rm train}^{(f)}$.

In mathematical formalism, a MFML model is given as:
\begin{equation}
    P_{\rm MFML}^{(F,\eta_F;f_b)}\left(\boldsymbol{X}_q\right) := \sum_{\boldsymbol{s}\in\mathcal{S}^{(F,\eta_F;f_b)}} \beta_{\boldsymbol{s}} P^{(\boldsymbol{s})}_{\rm KRR}\left(\boldsymbol{X}_q\right)~,
    \label{eq_MFML_linearsum}
\end{equation}
where, the linear combination is performed over the set of selected sub-models of MFML, $\mathcal{S}^{(F,\eta_F;f_b)}$. This selection is decided by the choice of the \textit{baseline fidelity}, $f_b$, and the number of training samples at the highest fidelity (also called the target fidelity), $N_{\rm train}^{(F)}=2^{\eta_F}$.
The sub-models for a MFML model with a given $F$, $\eta_F$, and $f_b$ are identified as discussed previously in ref.~\citenum{vinod_2024_oMFML}. 
In Eq.~\eqref{eq_MFML_linearsum}, $\beta_{\boldsymbol{s}}$ are the coefficients of the linear combination of these selected sub-models. 

For MFML, the $\beta_{\boldsymbol{s}}$ are selected as
\begin{equation}
    \beta_{\boldsymbol{s}}^{\rm MFML} = \begin{cases}
        +1, & \text{if } f+\eta_f = F+\eta_F\\
        -1, & \text{otherwise}
    \end{cases}~,
    \label{eq_MFML_beta_i}
\end{equation}
based on ref.~\citenum{zasp19a}. The value of $\boldsymbol{\beta}_s$ prescribed in the above equation can be better understood with the following example. Consider a 3 fidelity dataset, that is $f\in\{1,2,3\}$ with $f=3=F$, and $f=1=f_b=1$. If one uses $N_{\rm train}^{(3)}=2^2$ then $\eta_F=2$. Therefore, following Eq.~\eqref{eq_MFML_linearsum} and Eq.~\eqref{eq_MFML_beta_i}  one can build the MFML model 
$$P_{\rm MFML}^{(3,2;1)}(\boldsymbol{X}_q):= \left(P_{\rm KRR}^{(3,2)}(\boldsymbol{X}_q)-P_{\rm KRR}^{(2,2)}(\boldsymbol{X}_q)\right) + \left(P_{\rm KRR}^{(2,3)}(\boldsymbol{X}_q) - P_{\rm KRR}^{(1,3)}(\boldsymbol{X}_q)\right) + P_{\rm KRR}^{(1,4)}(\boldsymbol{X}_q)
~,$$ 
where the individual KRR models are built for the composite index, $\boldsymbol{s}$.  

Recently, a methodological development over conventional MFML was proposed and shown to be superior to MFML in ref.~\citenum{vinod_2024_oMFML}. Termed optimized MFML (o-MFML), it optimizes the combination of the sub-models of MFML. In other words, it considers coefficient values different from those defined in Eq.~\eqref{eq_MFML_beta_i} for the different sub-models by optimizing over a validation set $\mathcal{V}^F_{\rm val}:=\{(\boldsymbol{X}_q^{\rm val},y^{\rm val}_q)\}_{q=1}^{N_{\rm val}}$.
With some target fidelity, $F$ with $\eta_F$ given and a baseline fidelity $f_b$ chosen, one can formally define the o-MFML model as: 
\begin{equation}
    P_{\rm o-MFML}^{\left(F,\eta_F;f_b\right)}\left(\boldsymbol{X}_q\right) := 
    \sum_{\boldsymbol{s}\in \mathcal{S}^{(F,\eta_F;f_b)}}\beta_{\boldsymbol{s}}^{\rm opt} P^{(\boldsymbol{s})}_{\rm KRR} \left(\boldsymbol{X}_q\right)~,
    \label{eq_POM_def}
\end{equation}
where $\beta_{\boldsymbol{s}}^{\rm opt}$ are the optimized coefficients of MFML.
These values are attained by solving the optimization problem
\begin{equation}   
\beta_{\boldsymbol{s}}^{\rm opt} = \arg\min_{\beta_{\boldsymbol{s}}} 
    \left\lVert \sum_{v=1}^{N_{\rm val}} \left(y_v^{\rm val} - \sum_{\boldsymbol{s}\in S^{(F,\eta_F;f_b)}} \beta_{\boldsymbol{s}} P^{(\boldsymbol{s})}_{\rm KRR}\left(\boldsymbol{X}^{\rm val}_v\right)\right) \right\rVert_p~,
\end{equation}
where one minimizes some $p$-norm on the validation set defined above. 
The optimization procedure used in this work is ordinary least squares (OLS) which uses a $p=2$ norm. 

\subsection{Scaling Factors}\label{scalingfac}
\begin{figure}[htb!]
    \centering
    \includegraphics[width=0.8\linewidth, trim = 0cm 2.5cm 0cm 0cm, clip]{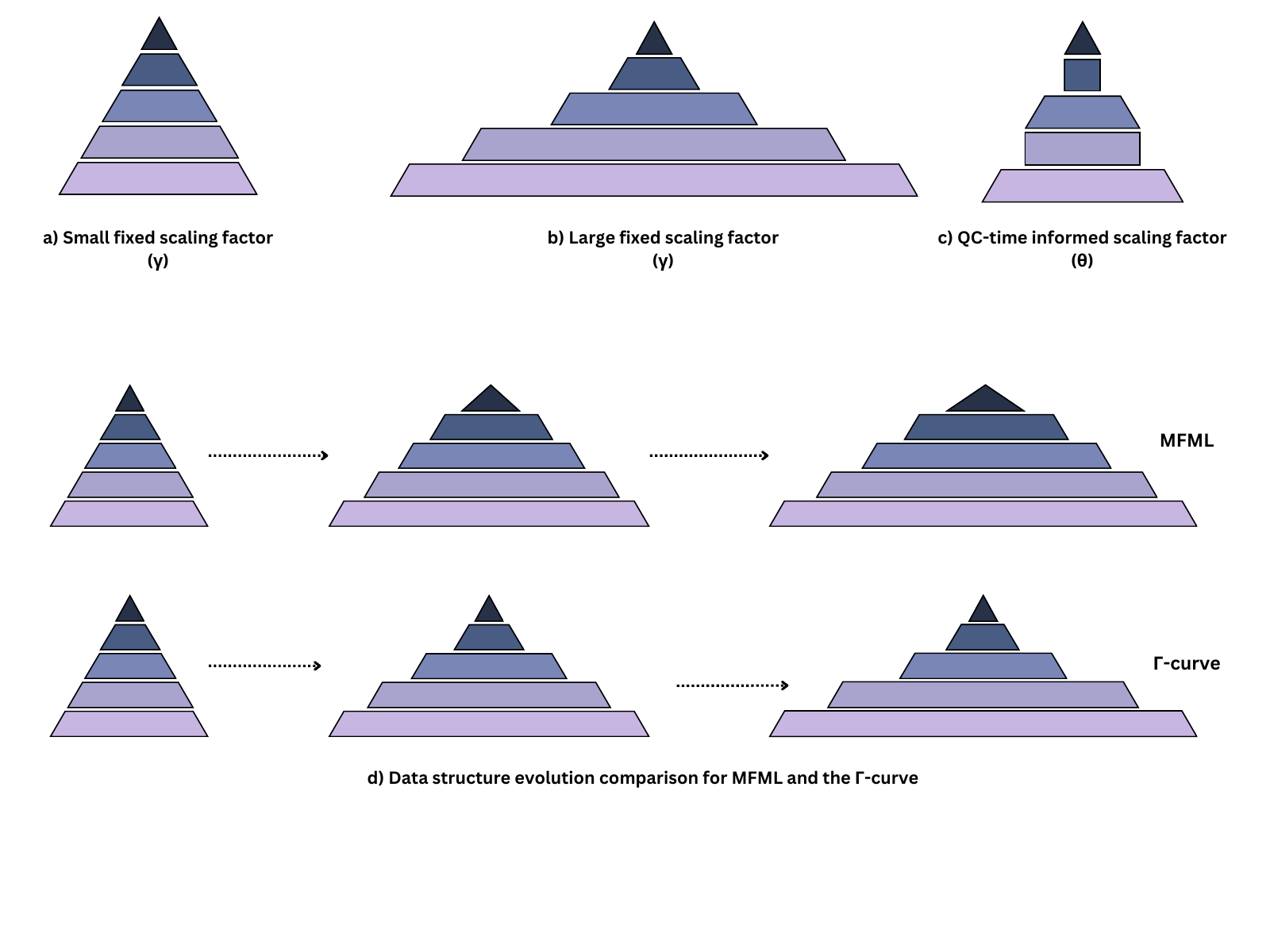}
    \caption{A hypothetical comparison of training data used across fidelities for the different kinds of scaling factors used in this work. a) The multifidelity training data structure used in MFML with a small fixed scaling factor ($\gamma$). b) Multifidelity training data structure for a large fixed scaling factor ($\gamma$) results in a larger number of training samples being used at the cheaper fidelities. 
    c) The structure of multifidelity training data used for scaling factors that are decided based on the QC-time cost, explained in \nameref{scalingfac} as $\theta_f^F$ and $\theta_{f-1}^f$. d) Comparison of training data structure evolution for conventional MFML and the $\Gamma$-curve introduced in \nameref{gammacurve_theory}. Notice how the number of training samples used at the target (the costliest) fidelity remain same across the data structure for the $\Gamma$-curve while they increase for the conventional MFML method.
    }
    \label{fig_scalingfactor_graphic}
\end{figure}
Thus far, in both MFML and o-MFML, the number of training samples used for each fidelity are scaled by a \textit{scaling factor} of $\gamma=$2 based on research on SGCT \cite{hegland2016combination,harbrecht2013combination,reisinger2013combination}. 
For example, if one has $N_{\rm train}^{(F)}=32$ training samples for fidelity $F$ then it is that $N_{\rm train}^{(F-1)}=64$ training samples for fidelity $F-1$, 
$N_{\rm train}^{(F-2)}=128$ training samples on fidelity $F-2$, and so on. 
The scaling up of the training samples as one decreases the fidelities can be thought of intuitively from a perspective of sparseness of data. As one increases the fidelity, the cost of QC calculation increases. This results in lower number of point-calculations that need to be made at this fidelity. 

The scaling factor can itself be varied to assess its effect on the model error. This work tests five such scaling factors, in particular, $\gamma \in\{2,3,4,5, 6\}$. For each of these scaling factors, the training set size increases exponentially as one goes down the fidelities. If one starts with $N_{\rm train}^{F}$ samples at the target fidelity, then at each lower fidelity, $f<F$, the number of training samples would be $N_{\rm train}^{f}= \gamma^{F-f} \times N_{\rm train}^{F}$. As an example, if $F=5$, $N_{\rm train}^{F}=2$, and $\gamma=3$, then for $f_b=1$ the training set size for each fidelity, in increasing order of the fidelity, would be $\{3^4\cdot 2,3^3\cdot 2, 3^2\cdot 2, 3\cdot 2, 2\}$. The variation of the training set sizes with increasing values of $\gamma$ is represented pictorially in Figure \ref{fig_scalingfactor_graphic}(a)-(b).

This work studies two additional approaches for QC-cost adapted selection of scaling factors. This approach takes into account the compute times for each fidelity before adaptively selecting the ratio of training samples between two consecutive fidelities.
Previous literature for MFML constructs these ML models with a scaling factor of 2 for each consecutive fidelity resulting from work on SGCT \cite{zasp19a}. Theoretical work in SGCT arrives at such a factor by analyzing the cost-benefit ratio of the different levels, or fidelities ~\cite{griebel1990combination, Bungartz_Griebel_2004, hegland2007combination}. 
Following this, the use of $\gamma=2$ in MFML is motivated by the assumption that the cost of fidelity $f$, that is the time to generate training data at $f$, versus that for $f+1$ differs by a factor of 2. Simple put, there is an inherent assumption that $T_{f+1}/T_{f}\approx2$. Therefore, one can consider the time-cost informed scaling factors between consecutive fidelities to be a generalization of this assumption.
The QeMFi dataset provides the QC compute time in seconds for each of the five fidelities when computed on a single core. 
This information can be used to determine a time-informed scaling factor for each fidelity as opposed to setting a single scaling factor for all the fidelities.

While there could be different ways to determine these time-informed scaling factors, the most trivial approach is to take the nearest integer value of the ratio of the compute times for the subsequent fidelity. That is, one can define $\theta_{f-1}^{f}:=\lfloor T^f/T^{f-1} \rceil$, where $\lfloor\cdot\rceil$ denotes integer rounding. 
This specific choice of scaling factors is made to take into account the relative time-cost of the fidelities used in the MFML model. It is reasonable to assume that the number of training samples used at consecutive fidelities should be based on the ratio of the cost of those fidelities.
Since QeMFi is a collection of different molecules, this approach was carried out with reference to the compute times for the largest molecule in the database: o-HBDI. This results in scaling factors $\theta_{f-1}^f=\{3,1,2,1\}$ for increasing fidelity. That is, at SVP, the same number of training samples as TZVP are used while at the 631-G fidelity, it is twice, and so on. However, MFML models are built in such a way that subsequently cheaper fidelities have some more training samples than the previous fidelity so that the difference between the sub-models can be taken. In order to achieve this, after the number of training samples are decided by the scaling factors, if fidelity $f-1$ has the same number of training samples as fidelity $f$, then one additional sample is added to the sub-model at fidelity $f-1$. As an example, if $N_{\rm train}^{\rm TZVP}=2$, then the training samples for the different fidelities would be $\{12,4+1,4,2+1,2\}$. Hereon, the MFML models built with this approach of scaling factors are referred by $\theta_{f-1}^{f}$.

A second approach of implicitly incorporating the time-cost of the fidelities is to take the ratio of the compute times with respect to the target fidelity. 
This approach is motivated by posing the question, what amount of training data used at a specific fidelity would cost the same as the training data used at the target fidelity.
Once again, the nearest integer value is considered. This leads to the definition of $\theta_{F}^{f}:=\lfloor T^F/T^f \rceil$ for all $f<F$. As for the case of $\theta_{f-1}^f$, the reference molecule was chosen to be o-HBDI. This leads to scaling factors $\theta_f^F=\{9,3,2,1\}$ for increasing fidelity. Since the SVP fidelity is scaled by a factors of 1, as discussed earlier, one additional training samples was added each time to maintain the multifidelity structure required for MFML. As an example, consider the case for $N_{\rm train}^{\rm TZVP}=2$. Then the training samples at the different fidelities would be $\{108,12,4,2+1,2\}$. This formulation of scaling factors is hereon associated with $\theta_{f}^{F}$. Scaling factors based on QC-compute cost as diagrammatically depicted in Figure \ref{fig_scalingfactor_graphic}(c).

\subsection{Error Contours of MFML}\label{method_errorcontour}
The prediction error of the different ML and MFML models assessed in this work are given as relative mean absolute errors (RMAE) which is calculated over the holdout test set as:
\begin{equation}
    \text{RMAE} = \frac{1}{N_{\rm test}}\sum_{q=1}^{N_{\rm test}}\left\lvert\frac{P_{\rm ML}\left(\boldsymbol{X}_q^{\text{ref}}\right) - {y}^{\text{ref}}_q}{y^{\text{ref}}_q}\right\rvert~.
    \label{eq_MAE}
\end{equation}
Due to the multifidelity data structure a simple cross validation approach cannot be used. Instead, the RMAE of the MFML models is calculated for a 10-run average as proposed and implemented in ref.~\citenum{vinod_2024_oMFML}, which accounts for the nested multifidelity data structure. For most ML approaches, the RMAE is reported for increasing training set sizes, thereby resulting in a learning curve. The learning curves is used as an indicator of the ability of the ML model to predict over unseen data.
Learning curves form a major part of the analysis offered in this work. In addition to the usual RMAE versus training samples learning curves, this work also studies the RMAE versus cost of generating training data for the multifidelity model as first proposed and implemented in ref.~\citenum{vinod23_MFML} 
for excitation energies.

The analysis of the learning curves for the different values of $\gamma$ (see  \nameref{LC_results} and  \nameref{time_cost_results}) indicate that the MFML training data structure needs only very little training samples at the higher fidelity. The contribution of each fidelity and the number of training samples at each fidelity is more complex than just the QC-time cost of the fidelity. In interest of studying the individual contribution of each fidelity and the diverse training samples choices at each fidelity, this work introduces a new error metric, namely, error contours. 
As a conceptual extension of learning curves, error contours of MFML report MFML model error as a function not simply of training samples chosen at a single fidelity but as a function of training samples, $N_{\rm train}^{f}$ and $N_{\rm train}^{f+1}$, chosen at two consecutive fidelities, $f$ and $f+1$. 
This form of analysis helps better understand the contribution of these fidelities to the overall model accuracy. Since the error contours can be studied for all consecutive fidelity pairs, these provide an in depth understanding of the contribution of each fidelity $f_b\leq f\leq F$ to the MFML model $P_{\rm MFML}^{(F,\eta_F;f_b)}$ in terms of model accuracy. Consecutive fidelities are studied instead of an arbitrary pair of fidelities in order to systematically assess what happens to the model error as one adds a cheaper fidelity while increasing the training set sizes.
In other words, the error contour is the RMAE of the MFML model by varying the training samples at two fidelities simultaneously. 
Since this work uses the QeMFi dataset that contains five fidelities, the error contours of MFML are studied for the following fidelity pairs: TZVP-SVP, SVP-631G, 631G-321G, and 321G-STO3G. Since the error contours are a function of two variables, $N_{\rm train}^f$ and $N_{\rm train}^{f+1}$, they are reported as contour plots.
Herein, error contours of MFML are discussed only for $\gamma=2$.  
The error contours give a better view into the contribution of a multifidelity data structure to model accuracy for a given target fidelity. 
The investigation of error contours for each fidelity pair indicates, in some sense, the weighted contribution of those fidelities to the overall model.  
A better understanding of this contribution will aid the choice of $N_{\rm train}^f$ for each fidelity that constitutes the MFML model.

\subsection{The \texorpdfstring{$\Gamma$}{Gamma}-Curve}\label{gammacurve_theory}
The study of the error contours in \nameref{results_error_contours} indicates that the multifidelity data structure can provide a high-accuracy model with a much lower number of costly training samples than the conventional MFML data structure approach. 
Coupled with the results of studying the learning curves for different scaling factors (see \nameref{time_cost_results} and \nameref{results_error_contours}), a new multifidelity approach is proposed: the $\Gamma$-curve.

In this approach, a fixed number of training samples are chosen at the highest fidelity, $N_{\rm train}^{\rm TZVP}$. An o-MFML model is built with $\gamma=2$. The cost of the training data is noted along with the model error over the holdout test set. For the next step of this curve, instead of varying $N_{\rm train}^{\rm TZVP}$, $\gamma$ is increased by an integer value. This curve is identified as $\Gamma(N_{\rm train}^{\rm TZVP})$-curve and is a measure of RMAE versus time-cost of training data of the multifidelity model for varying $\gamma$. For example, if one were to set $N_{\rm train}^{\rm TZVP}=2$, $f_b:$ 321G, then the $\Gamma(2)$-curve would be built with the first multifidelity training data structure (in increasing fidelity) as $\{2^3\cdot2,2^2\cdot2,2^1\cdot 2,2\}$. The next point would be built with a training data structure of $\{3^3\cdot2,3^2\cdot2,3^1\cdot2,2\}$ and so on. In general, for a $\Gamma(N_{\rm train}^{\rm TZVP})$-curve, the training data structure for $f:b$ 321G is given as $\{\gamma^3\cdot N_{\rm train}^{\rm TZVP},\gamma^2\cdot N_{\rm train}^{\rm TZVP}, \gamma^1\cdot N_{\rm train}^{\rm TZVP}\}$. For reasons explained in \nameref{time_cost_results} and \nameref{results_error_contours}, the STO3G baseline is not considered.
Herein, the $\Gamma(\cdot)$-curve is studied for $N_{\rm train}^{\rm TZVP}\in\{2,8,64\}$ to further assess the multifidelity structure of training data and evaluate the limits of the multifidelity approach. Since there is no trivial way to express the number of training samples used at a certain fidelity and relate it to the RMAE of the model, the $\Gamma(\cdot)$-curve is plotted only as RMAE versus multifidelity training data generation cost. The variation of the number of training samples used at each fidelity for the $\Gamma$-curve is contrasted with that for the learning curve of MFML in Figure \ref{fig_scalingfactor_graphic}(d). Notice the wider base of the pyramid for $\Gamma$-curve signifying a much larger number of training samples being used at the cheapest fidelities.   

\section{Results}\label{results}
This section presents the analysis of varying the scaling factor for MFML and o-MFML. The results are presented in two major formats. First, standard learning curves of RMAE versus number of training samples used at the highest fidelity of TZVP are presented.
Following this, the model error is assessed a function of the time-cost of generating the training data for the model. This assessment from ref.~\citenum{vinod23_MFML} informs of the effectiveness of the diverse models that are studied in this work.
Once these results are interpreted, error contours of MFML as described in \nameref{method_errorcontour} are studied.

\subsection{Learning Curves} \label{LC_results}
\begin{figure}[htb!]
    \centering
    \includegraphics[width=\linewidth]{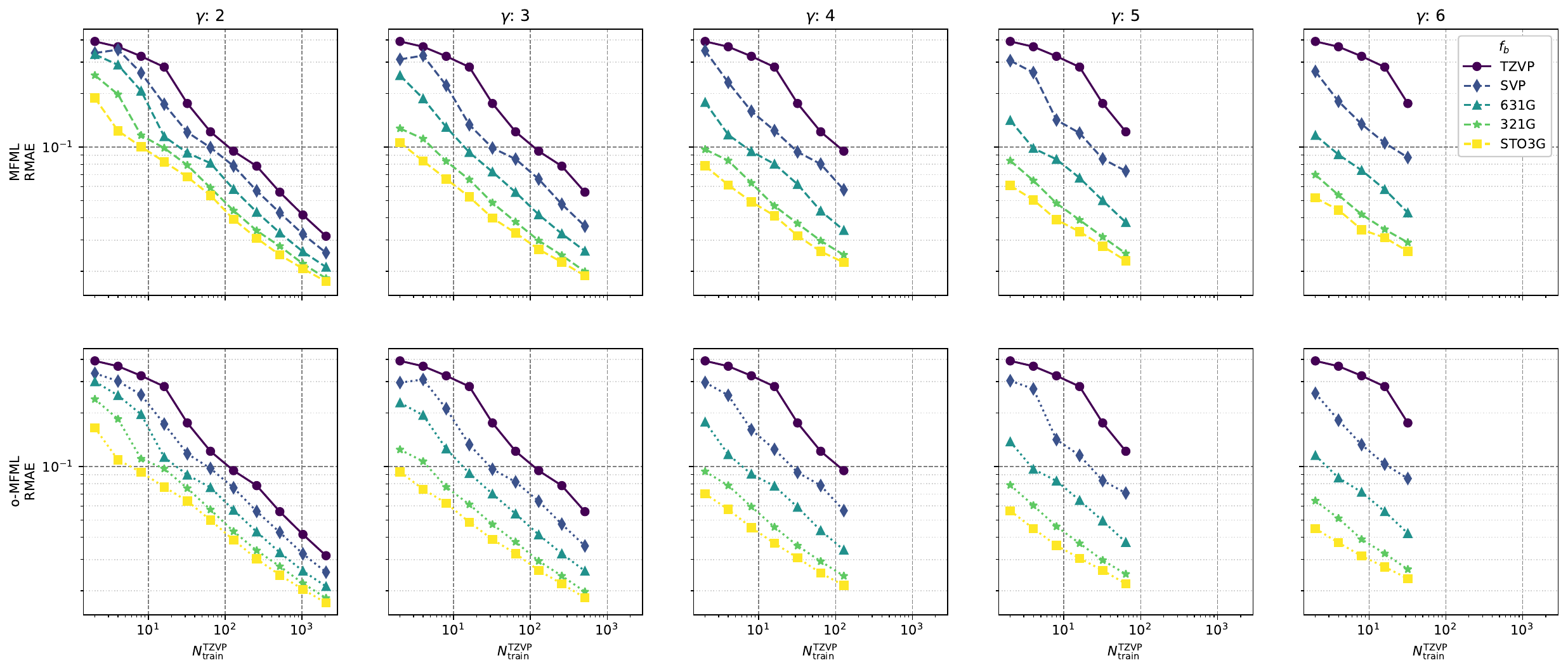}
    \caption{Multifidelity learning curves for the prediction of excitation energies taken from the QeMFi dataset. The top row corresponds to the MFML models while the bottom row is for the o-MFML models. Different fixed scaling factors are used to scale the data across each fidelity in the multifidelity models as explained \nameref{scalingfac}. The scaling factors are reported on the top of each column.}
    \label{fig_LC_EV}
\end{figure}

The primary assessment of the effect of different scaling factors is carried out using learning curves for the resulting MFML and o-MFML models. These learning curves are shown in Figure \ref{fig_LC_EV} for different scaling factors. In all cases, the scaling factors are constant across the different fidelities as explained in \nameref{scalingfac}. The top row of the figure depicts the learning curves for MFML while the bottom row corresponds to the o-MFML models. The learning curves are shown for varying baseline fidelities. A single fidelity KRR learning curve is also shown for reference. The RMAEs are reported as unitless quantities. 

In Figure \ref{fig_LC_EV}, the first column shows the learning curves for the scaling factor of 2. This is the original scaling factor used in refs.~\citenum{zasp19a, vinod23_MFML, vinod_2024_oMFML} and is used as a reference to evaluate the other scaling factors against. Within these reference results, one observes that the addition of cheaper baselines results in a constantly lowered offset of the learning curves. The interpretation from the lowered offsets is that similar models errors can be achieved with lower number of training samples at the target fidelity with the addition of cheaper fidelities. With the cheapest fidelity, STO3G, being added to the multifidelity model, one observes RMAE of 0.4 with around $200$ training samples at TZVP.
In comparison to this, an increase of of the scaling factor, $\gamma$, provides MFML models that achieve similar errors for lower number of training samples at TZVP. For example with a scaling factor of 3, the STO3G baseline MFML model achieves an RMAE of 0.4 with $N_{\rm train}^{\rm TZVP}=32$. With a scaling factor of 6, the number of training samples at TZVP needed to achieve this same error is lowered further to about 4. The learning curves for o-MFML also indicate the same across varying scaling factors. There is little difference between the learning curves for MFML and o-MFML. This could be due to the MFML combination being already optimal for this multifidelity data structure as has been argued in ref.~\citenum{vinod_2024_oMFML}. There is however, slight improvement in all cases of o-MFML and it does result in reduced RMAEs across the different scaling factors.

From these results it appears that a higher number of training samples with cheaper fidelities improves the predictive capabilities of the MFML and o-MFML models. One possible reason for this could be that the use of larger data at the cheaper fidelities results in more information about the overall multifidelity structure being included into the MFML models. 

\begin{figure}[htb!]
\centering
    \includegraphics[width=0.75\linewidth]{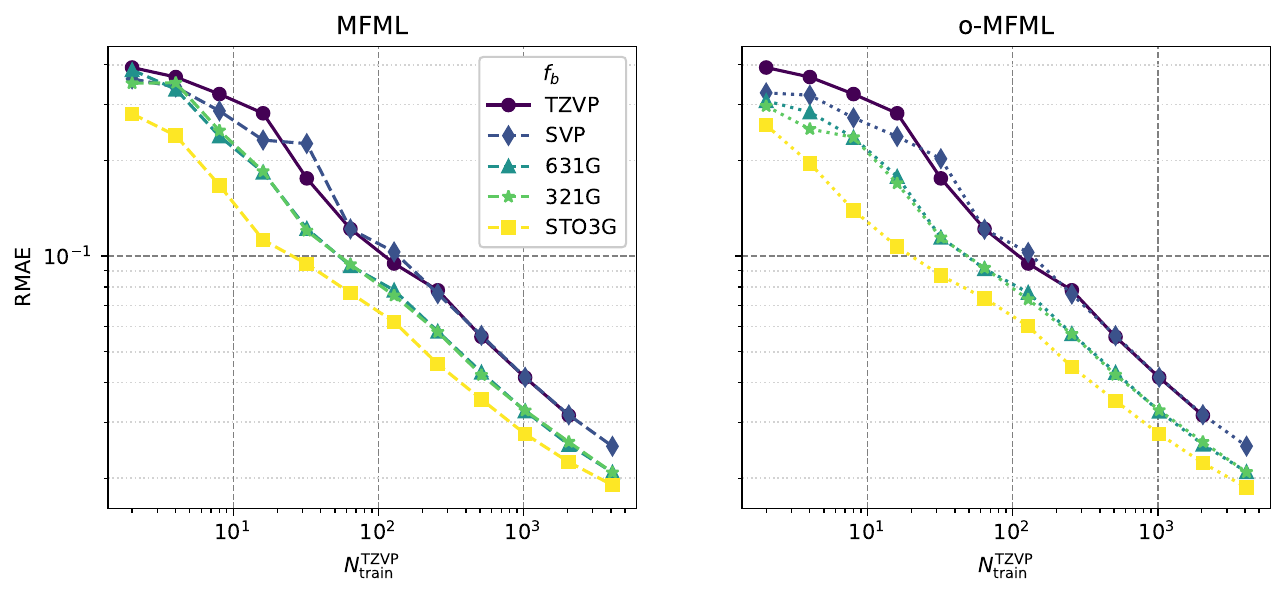}
    \caption{MFML and o-MFML learning curves for scaling factors, $\theta_{f-1}^f$, between fidelities chosen as ratios of the QC compute time of subsequent fidelities. Single fidelity KRR at TZVP is also shown for reference. Single fidelity KRR learning curves are also provided for reference. The legend describes the baseline fidelity, $f_b$, of the multifidelity model.}
    \label{fig_ffm1_LC}
\end{figure}
\begin{figure}
    \centering
    \includegraphics[width=0.75\linewidth]{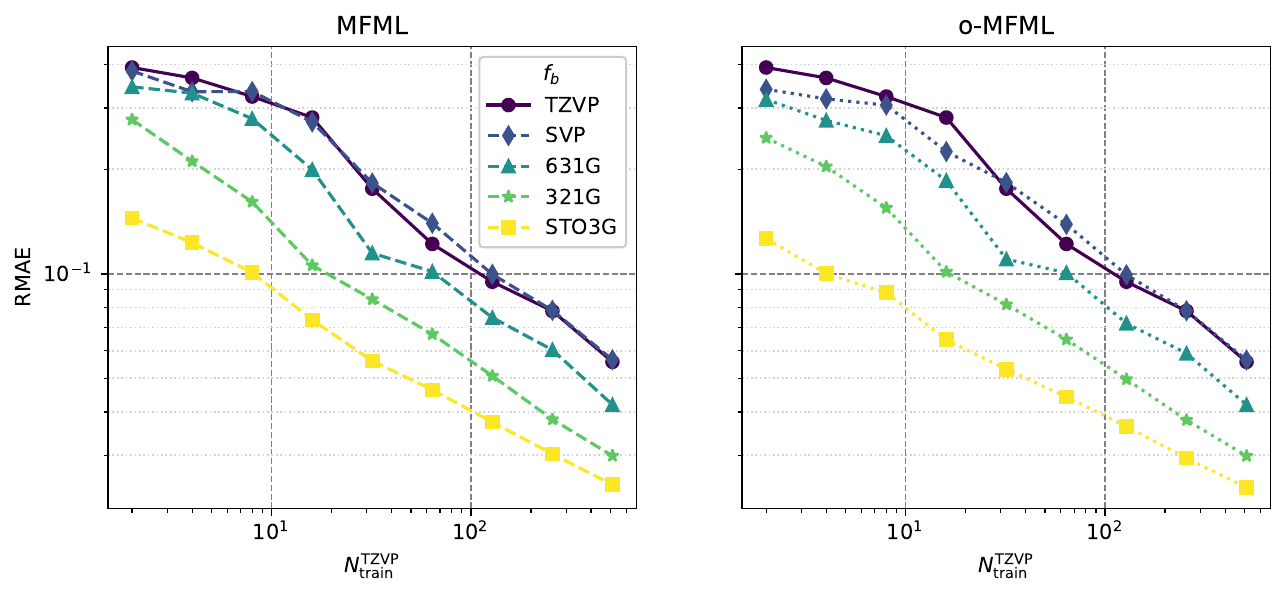}
    \caption{MFML and o-MFML learning curves for scaling factors, $\theta_f^F$, between fidelities selected as ratios of the QC compute time of that fidelity to the compute time of TZVP, that is the target fidelity. Single fidelity KRR learning curves are also provided for reference. The legend describes the baseline fidelity, $f_b$, of the multifidelity model.}
    \label{fig_target_fidelity_ratio}
\end{figure}

In addition to the fixed scaling factors across fidelities, two special cases of scaling factors were introduced in \nameref{scalingfac} based on the QC compute time of each fidelity. These were denoted by $\theta_{f-1}^f$ and $\theta_f^F$ as explained in \nameref{scalingfac} in detail. Learning curves were generated for both MFML and o-MFML models for both these cases. The results are shown in Figures \ref{fig_ffm1_LC} and \ref{fig_target_fidelity_ratio} for both scaling factor cases with various baseline fidelities. The single fidelity KRR learning curves is also depicted for reference. 

Figure \ref{fig_ffm1_LC} depicts the results for $\theta_{f-1}^f$ with the left pane for MFML and right pane for o-MFML. 
As explained in \nameref{scalingfac}, these scaling factors are based on the ratio of QC-compute times of subsequent fidelities. Between some fidelities - namely between TZVP and SVP, and between 631G and 321G - this scaling was observed to be $1$. It is anticipated that these fidelities will not significantly improve the MFML models since there is very little additional information that is being added to the model. 
Indeed, as seen in Figure \ref{fig_ffm1_LC}, the multifidelity model built with SVP baseline does not provide any improvement over the single fidelity KRR. This is due to the fact that the number of training samples at both fidelities are nearly identical, only different by 1 sample, due to the scaling factor. This same observation can be made for the learning curves with 321G as baseline fidelity. With 631G and STO3G baselines, however, one observes improvement of the MFML and o-MFML models. With the STO3G baseline, MFML and o-MFML reach an RMAE of 0.03 with roughly 500 training samples at TZVP. 

Similarly, Figure \ref{fig_target_fidelity_ratio} reports the learning curves for $\theta_f^F$. The left-pane shows the results for MFML, while the right pane shows those for o-MFML. 
The SVP baseline fidelity once again shows very little improvement over the single fidelity KRR due to the scaling factor being unity (see \nameref{scalingfac}). 
However, each additional cheaper baseline fidelity, results in lowered offsets of the corresponding learning curves. With $N_{\rm train}^{\rm TZVP}=256$, the multifidelity models with the STO3G baseline result in RMAE of $\sim$0.03. 

\begin{figure}[htb!]
    \centering
    \includegraphics[width=0.8\linewidth]{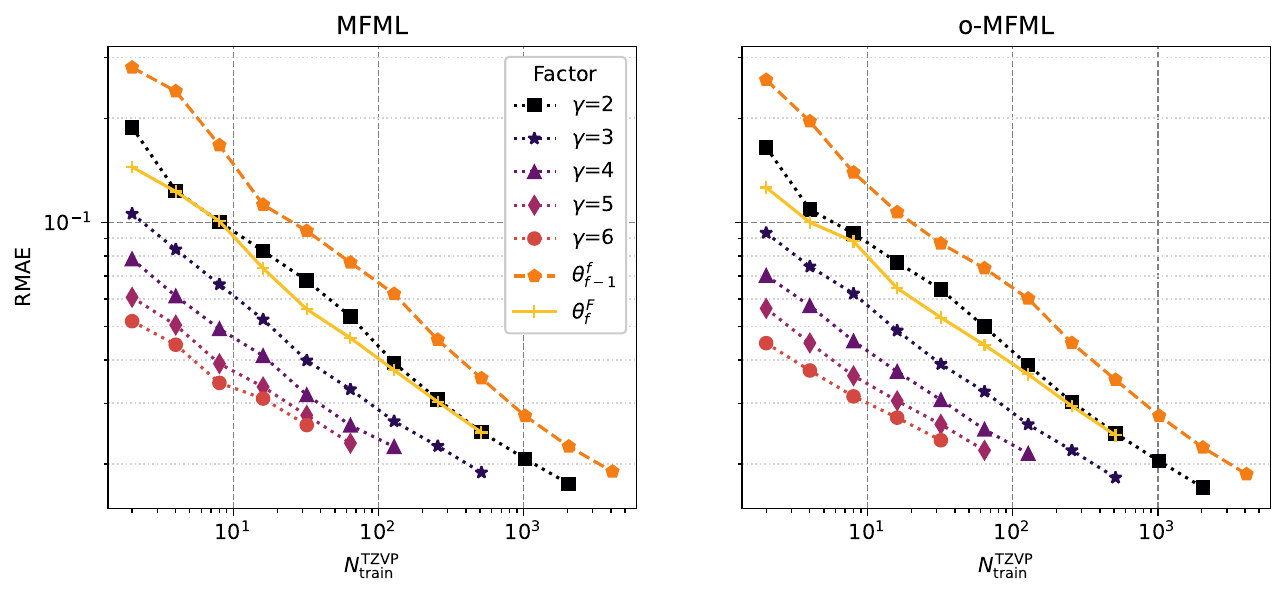}
    \caption{Comparison of learning curves for fixed scaling factors $\gamma$, $\theta_{f-1}^f$, and $\theta_f^F$ with $f_b$: STO3G. The x-axis reports the number of training samples used at the highest fidelity, that is, TZVP. Both MFML and o-MFML models are compared. Increasing values of $\gamma$ result in a constant lowered offset of the learning curves. The cost informed scaling factors show a higher value of MAE.}
    \label{fig_LC_comparisons_for_STO3G}
\end{figure}

\begin{table}[htb!]
    \centering
    \begin{tabular}{|c|c|c|}
    \hline
         \textbf{Factor} & \textbf{MFML} & \textbf{o-MFML} \\
         \hline
         \hline
         2 & 0.0790 & 0.0754 \\
         3 & 0.0486 & 0.0472 \\
         4 & 0.0371 & 0.0297 \\
         5 & 0.0312 & 0.0264 \\
         6 & 0.0290 & 0.0264 \\
         $\theta_{f-1}^f$ & 0.1201 & 0.1143 \\
         $\theta_{f}^F$ & 0.0844 & 0.0854 \\
         \hline
         
    \end{tabular}
    \caption{RMAE rounded off to 4 decimal points for MFML and o-MFML models built with the STO3G baseline fidelity for $N_{\rm train}^{\rm TZVP}=2^5$. This allows for a uniform comparison of the model accuracy not just between MFML and o-MFML but also across the scaling factors that are studied in this work. Notice that the learning curve for $\gamma=6$ only goes up to $N_{\rm train}^{\rm TZVP} = 2^5$ and therefore this is chosen as a comparison point for all other curves.}
    \label{tab_sto3g_error}
\end{table}

To aid comparison of the different scaling factors discussed so far, Figure \ref{fig_LC_comparisons_for_STO3G} depicts the learning curves for the MFML and o-MFML models built with the STO3G fidelity as baseline. The various factors are delineated in the legend of the plot. This plots shows that increasing values of $\gamma$ result in a lowered constant offset of the learning curves. In contrast, the multifidelity models built with time-informed scaling factors, $\theta_{f-1}^f$ and $\theta_f^F$ both show the highest model error.
This observation is consistent for both MFML and o-MFML models as can be seen from the two plots shown in Figure \ref{fig_LC_comparisons_for_STO3G}. 
Furthermore, the o-MFML models show lower errors than the MFML counterparts for all the cases 
as seen in table \ref{tab_sto3g_error} which reports the RMAEs for the MFML and o-MFML models with various scaling factors for the STO3G baseline for $N_{\rm train}^{\rm  TZVP}=2^5$ for ready reference. This training set size is chosen so that there is uniform comparison between the different scaling factors.
The behavior of the MFML models with increasing $\gamma$ is in some sense expected since an increasing value of the scaling factor implies an increased amount of training data, albeit only at the cheaper fidelities. This could be one potential reason to explain the lowered offsets that are observed. An increased amount of training samples at the lowered fidelities, due to the nested structure of the multifidelity training data, could impart meaningful information about the conformational phase-space and its relation to the excitation energies. 
The limited improvement that is seen from the learning curves of $\theta_{f-1}^f$ and $\theta_f^F$, which both had a much larger number of training samples at the cheapest fidelity in comparison to the other fidelities, could be due to the lack of sufficient training data in the fidelities that lie between the baseline and target fidelities. Furthermore, the value of $\theta_{f-1}^f$ for the SVP and 321G fidelities was $1$ which did not provide any additional information to the MFML model as was pointed out in the discussion for Figure \ref{fig_ffm1_LC}. This in turn affects the overall model that is built with the STO3G baseline fidelity. A similar argument can be made for why the MFML model with $\theta_f^F$ has limitations. Regardless, the MFML model built with $\theta_f^F$ does in fact achieve model RMAE that are comparable to the MFML model built with $\gamma=2$.

The results for fixed scaling factors, $\gamma$, indicate that a higher $\gamma$ results in a lower model error, or, smaller number of training samples at the costly target fidelity are needed. For the time-informed scaling factors, it was seen that these do not perform as well as was anticipated.  
However, one must be cautious about the results from Figure \ref{fig_LC_EV} and Figures \ref{fig_ffm1_LC} and \ref{fig_target_fidelity_ratio} before considering them to be improvements over the conventional MFML method with $\gamma=2$. Since one uses much more training samples at the cheaper fidelities as one increases the value of $\gamma$, the cost of generating training data needs to be assessed to better understand the cost-accuracy trade-off in these multifidelity models. In interest of such an analysis, the time-cost of generating training data versus model RMAE are discussed in the next section.  
Although the MFML and o-MFML models show similar RMAEs, only o-MFML models are discussed hereon. This is due to the observation of ref.~\citenum{vinod_2024_oMFML} that the o-MFML method provides a superior model even in cases of poor data distribution of the cheaper fidelities. That is, in the case of a poor data distribution, the o-MFML provides the better model in comparison to MFML.
Since all o-MFML models use the same validation set, the cost of generating the validation set is not included in the time-analysis plots. 

\subsection{Time Benefit Analysis}\label{time_cost_results}
\begin{figure}[htb!]
    \centering
    \includegraphics[width=\linewidth]{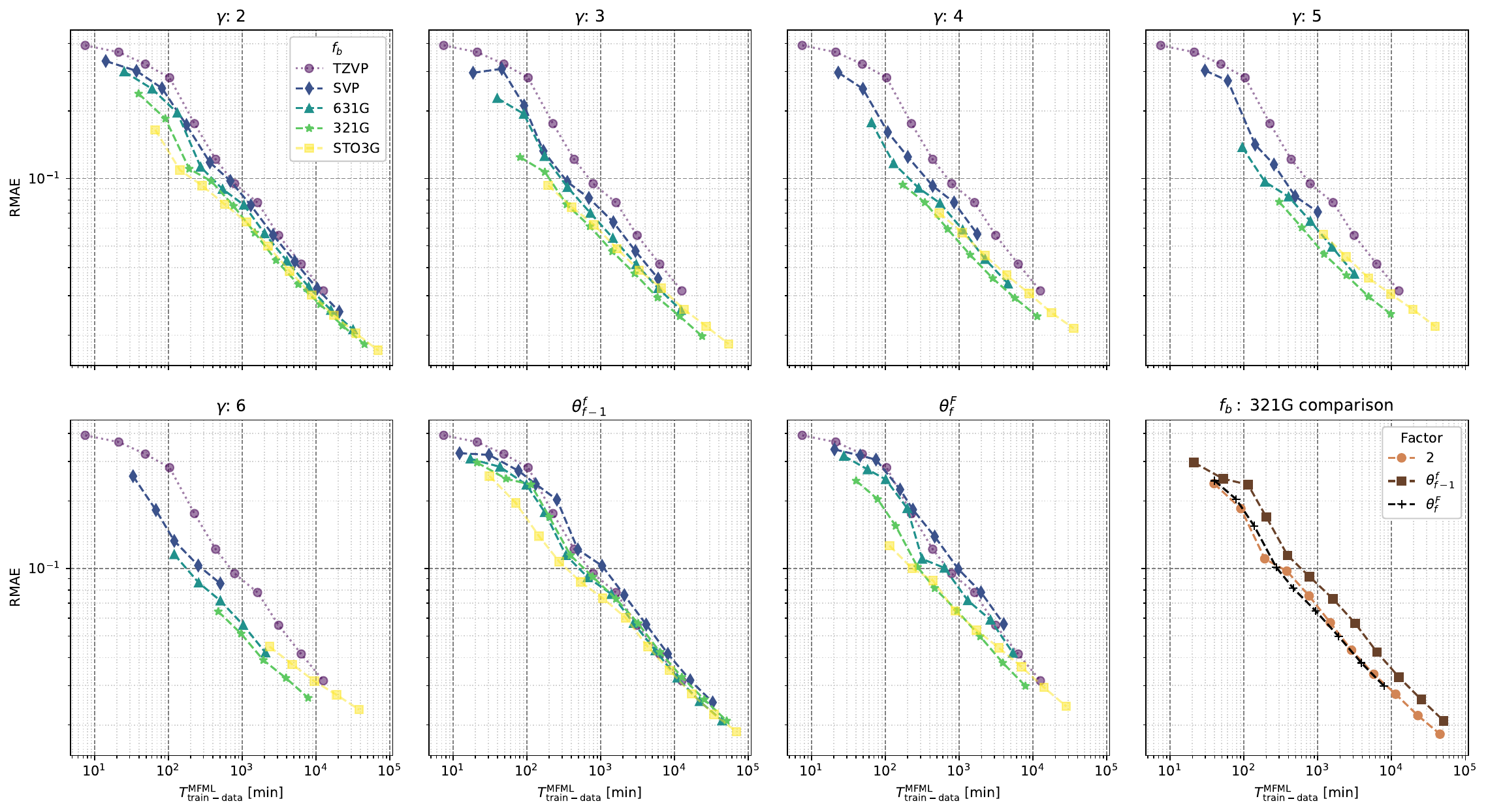}
    \caption{Time to generate training data versus RMAE of the corresponding o-MFML model for the diverse scaling factors studied. The different scaling factors used are denoted as sub-titles. The RMAE is unitless while the time-cost is in minutes. The single fidelity KRR case is also depicted for reference. As one increases the scaling factors across the fidelities, one observes that the learning curves of the MFML models shifts further due to the larger amount of training samples used. The two cases of $\theta_{f-1}^f$ and $\theta_f^F$ are explained in \nameref{scalingfac}. The bottom-right corner plot compares the o-MFML curves for the 321G baseline for the two time-informed scaling factors and the case of $\gamma=2$.}
    \label{fig_time_MAE_combined}
\end{figure}

As presented in refs.~\citenum{vinod23_MFML, vinod2024QeMFi_paper}, a good assessment of multifidelity methods is the study of model error versus the time to generate the training samples for the model. 
In interest of such a study, the RMAE versus training data generation time are studied for the just discussed test cases. The time to generate data for a multifidelity model is the sum over the times for generation of all the training samples used at all fidelities that form the multifidelity model. That is, $T_{\rm train-data}^{\rm MFML}:=\sum_{f_b\leq f\leq F} N_{\rm train}^f\cdot T_{QC}^f$ where $N_{\rm train}^f$ is the number of training samples used at some fidelity $f$, and $T_{QC}^f$ is the corresponding single-point QC-compute time for that fidelity. The QC-compute times recorded in the QeMFi dataset are those for a single-core computation and are provided for each fidelity for each molecule type \cite{vinod2024QeMFi_paper}.

The RMAE versus $T_{\rm train-data}^{\rm MFML}$ plots for the various scaling factors are shown in Figure \ref{fig_time_MAE_combined} for o-MFML. Only o-MFML is shown since it has a lower error compared to MFML for all the cases (see Figures \ref{fig_LC_EV},\ref{fig_ffm1_LC}, and \ref{fig_target_fidelity_ratio}). The RMAE and the time axes are both presented in log-scaled values. 
The axes of the plots are scaled identically for easy comparison among the different scaling factors. 
The bottom-right corner plot compares the time-cost based scaling factors to the case of $\gamma=2$ for the MFML model built with $f_b$: 321G and not for the cheapest STO3G baseline for reasons discussed below.

For the different cases of scaling factors shown in Figure \ref{fig_time_MAE_combined}, it can be seen that the addition of cheaper baselines helps achieve a specific model accuracy with less time cost to generate the training data. In general, fixing a specific MAE, one can see that the curves of the cheaper baseline achieve this error earlier with respect to the time axis. Alternatively, if one were to set a time budget and draw a vertical line at that value (as on the x-axis), then the cheaper baseline models result in lower RMAEs than the single fidelity KRR model. 
The case of STO3G baseline is an exception. For all scaling factors, the addition of the STO3G baseline does not provide significant improvement of the model. In fact, it increases the training data generation cost.
The STO3G energies do not provide major improvement to the o-MFML model over the 321G baseline. 
This could be due to poor data distribution that has previously been noted for the STO3G fidelity for excitation energies of molecules \cite{vinod23_MFML, vinod_2024_oMFML, vinod2024_nonnestedMFML}. The time-cost versus RMAE plots make this evident. Although the analysis of conventional learning curves from \nameref{LC_results} indicated that the STO3G baseline fidelity improved the MFML model, these time-cost plots indicate that this comes at a cost which supersedes the RMAE improvement that is observed.  
However, consider the case of the special scaling factors $\theta_f^F$, which are decided by the ratio of the QC-compute times of a fidelity $f$ to the QC-compute time of the target fidelity $F$. For some portions of learning curve for  $f_b=$ STO3G, o-MFML does provide lower errors as is expected from such multifidelity models \cite{vinod_2024_oMFML}. 
This could indicate that the use of o-MFML could improve the model accuracy even for the cases of poor data distribution as seen in the STO3G fidelity.
For the o-MFML models that are built with the 32G baseline fidelity, the time benefit of the multifidelity approach becomes all the more perceptible across the various scaling factors. For instance, in the case of $\gamma=3$, the o-MFML model results in an RMAE of 0.1 with a time cost of $\sim 200$ minutes. The KRR model achieves a similar error with a time-cost of $\sim 1,000$ minutes. This indicates a time benefit of about 5 times with this baseline fidelity for $\gamma=3$. Similarly, for $\gamma=5$, the KRR model achieves an RMAE of 0.07 with a time cost of $\sim 2,000$ minutes while the o-MFML model achieves a similar error for a time cost of $\sim 300$ minutes resulting in a time benefit of about 6 times. Similar observations can be made for the other values of $\gamma$.
The time benefit is less pronounced for the cases of $\theta_{f-1}^f$ and $\theta_f^F$ but is still present for $f_b:$ 321G. 

While each scaling factor does improve the time-cost needed to achieve a certain RMAE vis-\'a-vis the single fidelity KRR, it is also important to see which scaling factor performs better with respect to the others for a given baseline fidelity. 
The bottom-right plot of Figure \ref{fig_time_MAE_combined} compares the time-cost versus RMAE curves of MFML models for $\gamma=2$, $\theta_{f-1}^f$, and $\theta_f^F$ for the baseline fidelity of 321G. The STO3G baseline is not considered due to its poor distribution. These specific scaling factors are chosen to better understand the standing of the time-informed scaling factors with respect to the fixed scaling factors. 
This comparison in Figure \ref{fig_time_MAE_combined} for the 321G fidelity shows that the fixed scaling factor of $\gamma=2$ performs better than both the time-cost informed scaling factors (see \nameref{scalingfac}). The MFML model built with $\theta_f^F$ does perform only as well as that built with $\gamma=2$, which is the default set-up for MFML. 
This could indicate that just the QC-compute time-cost information might not suffice to select the training samples at each fidelity. It could be that the model accuracy and multifidelity training structure relation is more complex than just accounting for the QC-compute cost. 
To better understand how each fidelity and the number of training samples at each fidelity contribute to the overall model error, the next section studies a new error metric, error contours (see \nameref{method_errorcontour} for details). This is intended to give a better view into the inner mechanisms of the multifidelity data structure in building a MFML model.

\subsection{Multifidelity Error Contours} \label{results_error_contours}
\begin{figure}
    \centering
    \includegraphics[width=\linewidth]{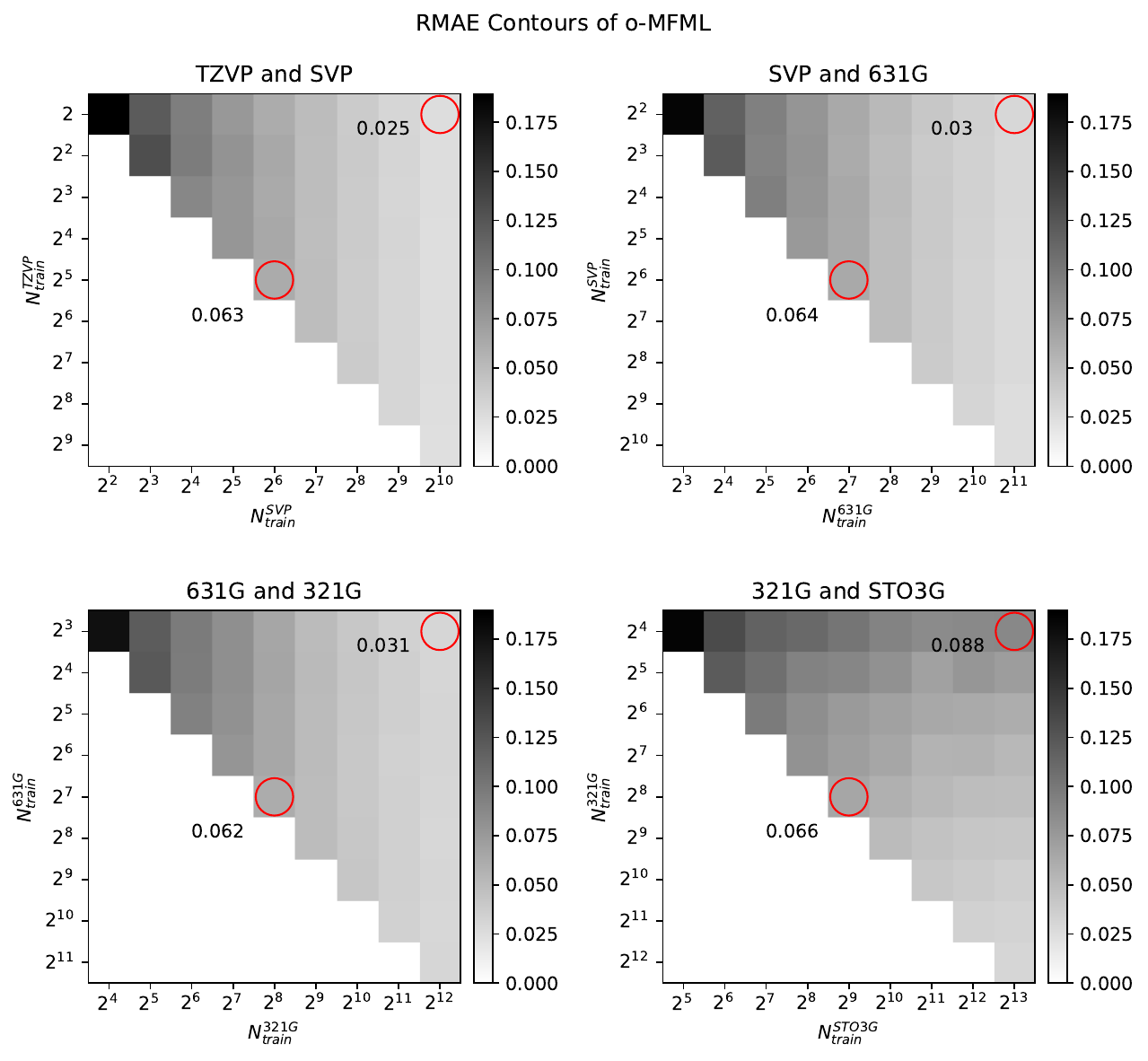}
    \caption{Excitation energy prediction errors with o-MFML for different training samples at different fidelities. The details of the method are explained in \nameref{scalingfac} for each case.
    In each plot, the vertical axis depicts the number of training samples used at the costlier fidelity, $f$, while horizontal axis reports the training samples used at the cheaper fidelity $f-1$. The resulting error for the o-MFML model with the specific choice of training samples used at fidelity $f$ and $f-1$ are depicted as the error contours.
    Here, the RMAE are depicted as contour plots for different training samples spanned across two fidelities. Two specific RMAEs are enumerated for all 4 cases: first, that for the smallest training set size at the higher fidelity, $f$, and the largest training set size at $f-1$; second, for the case where the training sample at $f$ and $f-1$ have the scaling factor of 2.}
    \label{fig_error_contours}
\end{figure}
The time versus RMAE results for different scaling factors hint that one might not necessarily need many training samples at the target fidelity. This would imply that one could build a cheap multifidelity model with a large number of training samples at the cheaper fidelities and then `raise' it to the target fidelity with an exceptionally small number of training samples at the target fidelity. This can be further studied with the error contours of multifidelity. These contours involve studying the model prediction error by varying the training sizes along fidelity $f$ and $f-1$ for all $f\leq F$. As was discussed in \nameref{scalingfac}, this is performed for consecutive fidelity pairs TZVP-SVP, SVP-631G, 631G-321G, and 321G-STO3G.

Figure \ref{fig_error_contours} illustrates the multifidelity error contours of o-MFML for different fidelity pairs for $\gamma=2$.
Consider the top-left plot corresponding to the TZVP-SVP fidelity pair. The y-axis denotes the number of training samples used at TZVP, while the x-axis depicts the number of training samples used at the SVP fidelity. The colors of the plot itself correspond to the MAE. 
In the usual o-MFML approach, the number of training samples used at SVP with respect to the number of training samples used at TZVP would be scaled by the factor $\gamma$ (in this case by 2). However, for this set-up there is no trivial scaling of training data that is carried out. Instead, the multifidelity model is built with a specific selection of training samples. For example, take the case for $N_{\rm train}^{\rm TZVP}=2$ and $N_{\rm train}^{\rm SVP}=2^{10}$ which is marked by the top-corner red circle. The RMAE reported here, 0.025, is for a multifidelity model that is built with the following multifidelity training structure (with increasing fidelity): $\{2^3\cdot2^{10}, 2^2\cdot2^{10}, 2\cdot2^{10},2^{10},2\}$. In other words, the scaling factor is only applied for the fidelities that are not studied as part of the error contour.
In contrast, in the usual o-MFML the training data structure would be $\{2^4\cdot2,2^3\cdot2,2^2\cdot2,2\cdot2,2\}$, this is the block that corresponds to $(N_{\rm train}^{\rm SVP}=2^2,N_{\rm train}^{\rm TZVP}=2)$ on the plot. The accompanying color-bar depicts that this regular o-MFML model results in a higher RMAE than 0.025. In general, the diagonal of the contour plot depicts the regular o-MFML model which is identified in the learning curves of Figure \ref{fig_LC_EV} for $\gamma=2$. The RMAE for $(N_{\rm train}^{\rm SVP}=2^6,N_{\rm train}^{\rm TZVP}=2^5)$ is highlighted as well reporting an RMAE of about 0.063 which is over twice of what is observed for $(N_{\rm train}^{\rm SVP}=2^{10},N_{\rm train}^{\rm TZVP}=2)$. This is a remarkable observation in that simply using two training samples at TZVP while increasing the training size at the lower fidelities results in a model that is more than twice as accurate. 
Furthermore, the RMAE for the block $(N_{\rm train}^{\rm SVP}=2^{10},N_{\rm train}^{\rm TZVP}=2)$ is similar to the model block $(N_{\rm train}^{\rm SVP}=2^{10},N_{\rm train}^{\rm TZVP}=2^9)$. In general, it is seen that a lower number of TZVP training samples with a larger training set size at the cheaper fidelities results in more accurate multifidelity model. 

Similar observations and inferences can be made for the error contour for the SVP-631G fidelity pair as seen on the top-right plot of Figure \ref{fig_error_contours}. In this set-up, consider the top right corner which is marked with a circle. This is identified as $(N_{\rm train}^{631G}=2^{11},N_{\rm train}^{\rm SVP}=2^2)$ and has the following multifidelity data structure (with increasing fidelity): $\{2^{13},2^{12},2^{11},2^2,2\}$. As in the previous case, the training data scaling is only applied to the fidelities that are not studied as part of the error contour. 
This mode reports 0.030 as the RMAE. Once again, the diagonal of the contour plot corresponds to the regular o-MFML model. Consider then, the block identified by $(N_{\rm train}^{631G}=2^{6},N_{\rm train}^{\rm SVP}=2^7)$ which has a training data structure (in increasing order of fidelity): $\{2^9,2^8,2^7,2^6,2^5\}$. This regular o-MFML model reports an RMAE 0.064, over twice as much as for the previous one. 
The overall contour plot reveals that the use of very few training samples at SVP paired with a larger number of training samples at the lower fidelities results in RMAEs that are comparable to the cases where one would use a lot more training samples at the SVP fidelity. In particular, this form of \textit{flattening} out the multifidelity training structure by using few training samples at the top fidelities and increasing the training samples at the cheaper fidelities, outperforms the regular o-MFML model (which are the diagonal blocks of the error contour).

Similar observations are made for the 631G-321G and 321G-STO3G pairs of fidelities which are seen in the bottom row of Figure \ref{fig_error_contours}. It is interesting to note, however, that the 321G-STO3G error contours do not follow the same trend as the others. Using very little 321G training samples and increasing the training samples at the STO3G fidelity does not result in lower RMAE as seen from the top-corner red marker error being 0.088 while the center marker reporting RMAE of 0.066. This is once again explained by the poor data distribution that has previously been reported for the STO3G fidelity \cite{vinod23_MFML, vinod_2024_oMFML, vinod2024_nonnestedMFML}.

The error contours for the multifidelity model hint at an interesting mechanism in the MFML approach. Based on the behavior of the model error as discussed above, it appears that one does not necessarily need to use many training samples at the higher fidelities, in particular at the target fidelity. 
This is indeed something that has been previously been hinted at in ref.~\citenum{dral2020hierarchical} using the optimization procedure for h-ML albeit with a larger number of training samples at the target fidelity. However, a thorough investigation of the multifidelity structure such as that performed in Fig.\ref{fig_error_contours} reveals that not only can a multifidelity model be built with low number of training samples at the costlier fidelities, but that this number is far smaller than what would be anticipated in the general MFML and similar methods. The error contours indicate that there is still a great deal of information available at the cheaper fidelities which only need to be `raised up' to the target fidelity with a surprisingly small number of training samples.
With such an understanding of the multifidelity training structure, one can begin to think of ways to select training samples at the different fidelities that need not necessarily follow the concept of a scaling factor between the fidelities. Furthermore, the results of varying $\gamma$ from Figure \ref{fig_time_MAE_combined} hint at a possible approach which is detailed in the following section.

\subsection{\texorpdfstring{$\Gamma$}{Gamma}-curves} \label{gammacurves_results}
The contour plots of Figure \ref{fig_error_contours} provide an interesting observation about MFML. One can potentially build cheaper multifidelity models by limiting the training samples used at the expensive fidelities and then proceeding to add cheap fidelity data to the multifidelity model. 

\begin{figure}[htb!]
    \centering
    \includegraphics[width=0.8\linewidth]{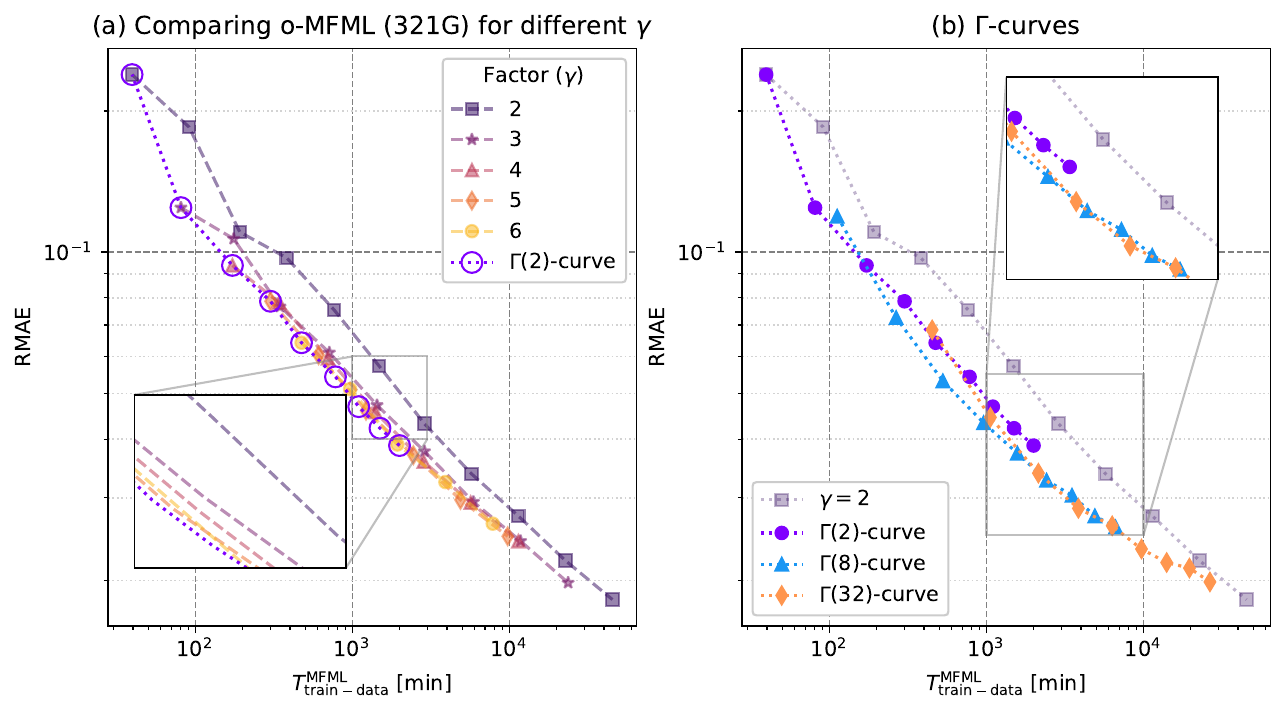}
    \caption{(a) Time to generate training data and corresponding o-MFML model error as RMAE for constant scaling factors, $\gamma$ used in this study. An inset between 1,500-3,000 minutes is provided for the comparison of the curves for all $\gamma$ studied in this work to readily compare in regions that are too crowded to be observed in the main plot. (b) RMAE versus time-cost for different $\Gamma(N_{\rm train}^{\rm TZVP})$-curves. Increasing the number of training samples at TZVP improves the model accuracies along the $\Gamma(\cdot)$-curves with a saturation observed towards the end of each curve.}
    \label{fig_adhoc_MFML_LC}
\end{figure}

Consider first the left-hand side plot of Figure \ref{fig_adhoc_MFML_LC} which shows the RMAE curves of o-MFML with $f_b:$ 321G for the different $\gamma$ studied in this work for comparison. An inset plot is provided which zooms into the region between 1,000-3,000 minutes to show the different curves clearly.  
In addition, a new curve as introduced in  \nameref{gammacurve_theory}, the $\Gamma(2)$-curve is depicted in the plot. The $\Gamma(2)$-curve is essentially the case where the number of training samples at TZVP are constrained to 2 but the remaining multifidelity data structure is allowed to grow as per the scaling factor $\gamma$.
That is, the $\Gamma(2)$-curve is built with the first point of the curves for the different $\gamma$ values. In some sense, this translates into it being a learning curve not as a function of $N_{\rm train}^{\rm TZVP}$ but rather of $\gamma$. 
From Figure \ref{fig_adhoc_MFML_LC}, it becomes evident that even for as little as 2 training samples at the highest fidelity, if one adds cheaper data to the multifidelity model - which corresponds to increasing the value of $\gamma$ without increasing $N_{\rm train}^{\rm TZVP}$ - the error of the o-MFML model decreases. 
For the same time-cost of a conventional o-MFML model built with $\gamma=2$, if one were to chose the models along the $\Gamma(2)$-curve, a lower RMAE can be achieved. The $\Gamma(2)$-curve in Figure \ref{fig_adhoc_MFML_LC} shows data points for up to $\gamma=10$ where the multifidelity training data structure (for increasing fidelity) is: $\{20000,2000,200,20,2\}$.
In the inset of the plot, one observes that the $\Gamma(2)$-curve results in errors that are lower than the o-MFML learning curves for fixed $\gamma$. However, the $\Gamma(2)$-curve converges to the o-MFML model built with $\gamma=6$. One potential reason for this could be the saturation of the multifidelity model built along the $\Gamma(2)$-curve. Due to a very large number of training samples at the cheaper fidelities (for instance, $2\cdot10^4$ at 321G for the last point on the $\Gamma(2)$-curve), the model is no longer able to clearly learn the correction between SVP and TZVP. 

The right-hand plot of Figure \ref{fig_adhoc_MFML_LC} further investigates this saturation by comparing the $\Gamma(N_{\rm train}^{\rm TZVP})$-curves for $N_{\rm train}^{\rm TZVP}\in\{2,8,32\}$. The y-axis reports the RMAE while the x-axis denotes the time-cost in minutes to generate the training data used in the multifidelity models.
An inset is provided for the interval between $10^3-10^4$ minutes for a better view of the $\Gamma(\cdot)$curves. The o-MFML learning curve for $\gamma=2$ is provided for reference.

Since the model errors throughout this work were reported in unitless RMAE, in order to understand how this translates to actual energy prediction, predictions are made for the holdout test set. For this purpose, the multifidelity model corresponding to $\Gamma(32)$ is used with $\gamma=10$. Using this model, the prediction of the first vertical excitation energies is made on the holdout test set. The absolute error values are then computed as $\lvert y_{\mathrm{ref}} - y_{\mathrm{pred}}\rvert$. The resulting values are reported in Table \ref{tab_errors}. This includes the mean of the absolute error, minimum absolute error, maximum absolute error, and the standard deviation of the absolute error. For all molecules, it can be seen that the model error is nearly identical indicating that the final multifidelity model built with the $\Gamma$-curve is not affected by the difference of the molecule. This is of course due to the fact that the training data consists of these molecules. In all cases, the model reports a mean absolute error close to 1 kcal/mol. 

\begin{table}[htb!]
    \centering
    \begin{tabular}{|c|c|c|c|c|}
    \hline
         \textbf{Molecule}& \textbf{mean} & \textbf{min} & \textbf{max} & \textbf{Std.~Deviation}\\
         \hline
         \hline
         \textbf{urea} & 1.0178 & 0.9940 & 1.0264 & 0.0035 \\
        \textbf{acrolein} & 1.0145 & 0.9928 & 1.0220 & 0.0034 \\
        \textbf{alanine} & 1.0145 & 0.9928 & 1.0222 & 0.0035 \\
        \textbf{SMA} & 1.0147 & 0.9928 & 1.0224 & 0.0035 \\
        \textbf{2-nitrophenol} & 1.0147 & 0.9928 & 1.0224 & 0.0035 \\
        \textbf{urocanic} & 1.0146 & 0.9928 & 1.0224 & 0.0035 \\
        \textbf{DMABN} & 1.0145 & 0.9928 & 1.0220 & 0.0034 \\
        \textbf{thymine} & 1.0145 & 0.9928 & 1.0224 & 0.0035 \\
        \textbf{o-HBDI} & 1.0146 & 0.9928 & 1.0224 & 0.0035 \\
        \hline
    \end{tabular}
    \caption{Absolute error values in kcal/mol between prediction and reference of excitation energies for the nine different molecules using $\Gamma$(32) for $\gamma=10$. The mean, range, and standard deviation of the absolute differences are presented.}
    \label{tab_errors}
\end{table}

\subsection{Transferability Assessment}
Transferability in ML refers to the concept of ML models being trained on specific type of dataset and having it predict the QC property for an out of sample dataset. For example, one could train ML models on the QeMFi dataset for excitation energies and use these trained ML models to predict the excitation energies of molecules that do not belong to the 9 molecules of the QeMFi dataset.
The transferability of ML models in QC has long been a challenging issue \cite{westermayr_2020_tranferML_UVabsoroption, Westermayr2020review, Zhang_2020_NN_tensorial_transfer}. 
In general, since ML is a statistical method, transferability is restricted by the type of data the model is trained on. That is, if a model is trained only on, say, benzene configurations, it is not expected to perform well in predicting for methanol or acrolein. 
Not only so, the type of molecular representation that is used in the model makes a major difference to the overall transferability of the model \cite{Westermayr2020review,Zhang_2020_NN_tensorial_transfer}.
However, in this subsection, the robustness of the $\Gamma$-curve method is investigated for transferability in order to complete the discussion on the development of this method. 

The QUESTDB database is a collection of several small molecules for which high accuracy excitation energies are available \cite{veril_2021_QUESTDB,loos_QUEST_2018}. The energies for these molecules are computed with mostly CC levels of theory. In order to properly assess the transferability, which is challenging as is, the excitation energies of 90 molecules from QUESTDB were computed with the DFT method using CAMB3LYP functional with the def2-TZVP basis set in the exact same manner as was done for the QeMFi dataset \cite{vinod2024QeMFi_paper}. 
Therefore, the predictions made on the molecules from QUESTDB are compared to the correct reference fidelity.
The error in prediction is studied as relative error (RE) which are computed as 
\begin{equation}
    \text{RE} := \left\lvert\frac{P_{\rm ML}\left(\boldsymbol{X}_q\right)-y_q^{\rm ref}}{y_q^{\rm ref}}\right\rvert
\end{equation}

The geometries from the QUESTDB database have an additional challenge in testing for transferability. This is the issue of index invariance in generating the unsorted CM molecular descriptor \citenum{david2020moleculardescriptors}. Since the geometries of the molecules from QeMFi are arranged in such a way to ensure index permutation invariance, one can use unsorted CM to train and evaluate ML models. However, this is not the case with the QUESTDB database. To overcome this fundamental issue, the following tests are performed using row-norm sorted CM representations  wherein the regular CM representation is built and then the rows are sorted based on $L_2$ norm\cite{Rup12CM, david2020moleculardescriptors}. That is, all models are trained and tested using sorted CM representations unlike the unsorted CM used in the preceding sections.
The protocol followed to demonstrate the performance of the $\Gamma$-MFML models is as given below:
\begin{itemize}
    \item Calculate the TD-DFT CAM-B3LYP def2-TZVP energies for the 90 molecules from QUESTDB
    \item Generate row-norm sorted CM for QeMFi
    \item Generate row-norm sorted CM for 90 molecules from QUESTDB
    \item Train single fidelity KRR model on QeMFi with $2^{10}$ training samples, Mat\'ern kernel from Eq.~\eqref{eq_matern} using $\sigma=200$, and $\lambda=10^{-10}$
    \item Train $\Gamma(8)$ and $\Gamma(32)$ MFML models with $\gamma=10$ on QeMFi as discussed in preceding sections
    \item Predict energies for the 90 molecules from QUESTDB using KRR and $\Gamma$-MFML models from step 3 and step 4
    \item Compare prediction to reference computed excitation energies from step 1
\end{itemize}

\begin{figure}[htb!]
    \centering
    \includegraphics[width=\linewidth]{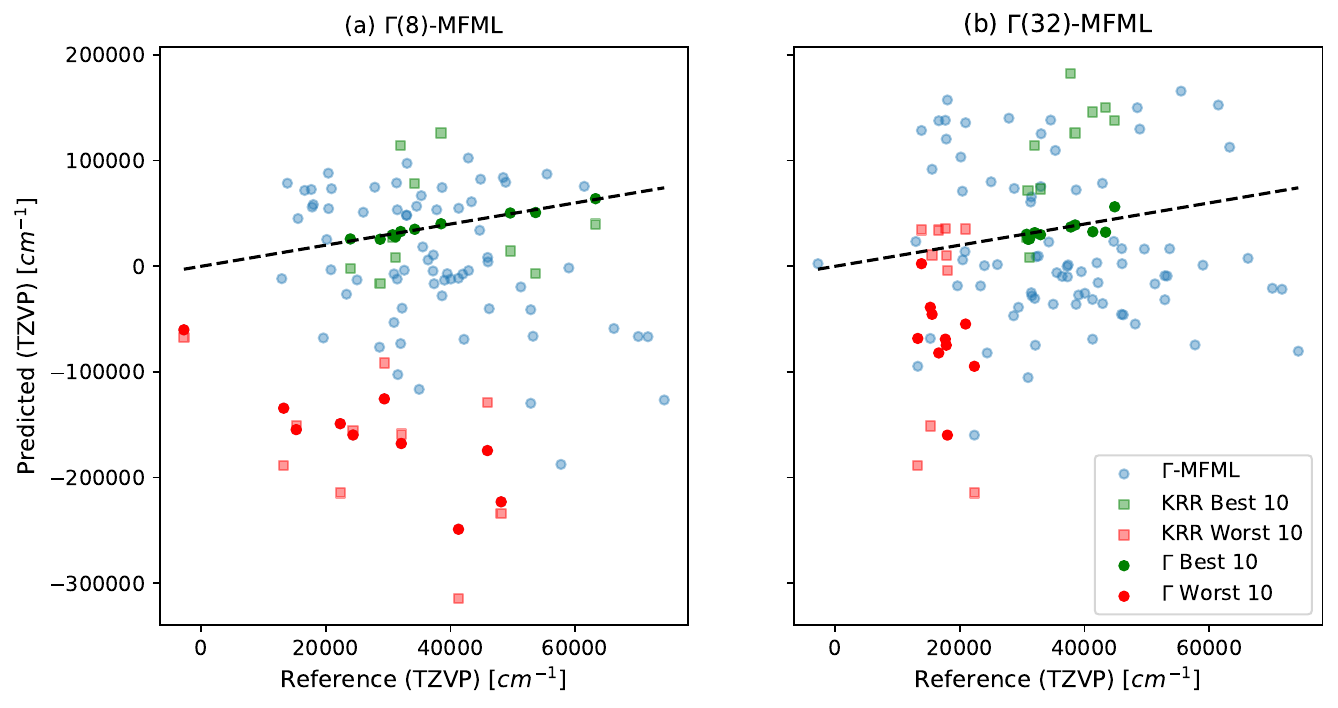}
    \caption{Scatter plot of reference and ML predicted excitation energies for transferability tests of $\Gamma$-curve on molecules from the QUESTDB database. The best 10 and worst 10 predictions are highlighted along with their predictions made using the single fidelity KRR model.}
    \label{fig_sortCM_questdb}
\end{figure}

A scatter plot of reference excitation energies versus ML predicted excitation energies is shown in Figure \ref{fig_sortCM_questdb}. The scatter plot is shown for $\Gamma(8)$ and $\Gamma(32)$ MFML models. The markers labeled in the legend as $\Gamma$-MFML correspond to the prediction versus reference of all 90 molecules chosen from QUESTDB.
The best 10, that is those with the lowest RE, are highlighted in green along with the worst 10 in red. For each of the $\Gamma$-curve models, for the best and worst predictions, the corresponding predictions of the single fidelity KRR model are also presented. This is done for two reasons. Firstly to indicate that it is a challenge even for single fidelity ML models to handle transferability tests. Secondly, it is interesting to assess how well the MFML models performed with respect to the single fidelity KRR models. 

Consider the left pane of Figure \ref{fig_sortCM_questdb} for $\Gamma(8)$ MFML model. The overall prediction of the $\Gamma(8)$ MFML model is poor for the QUESTDB database. There is a wide scatter of the points across the identity line (dashed black line). Admittedly, the best 10 predictions do lie on or close to this identity line. Consider the predictions made by the single fidelity KRR model for these very geometries which corresponds to the translucent green square markers. These are much further away from the identity line. In other words, the $\Gamma$-curve method is somewhat better than the single fidelity KRR method. Even for the 10 worst predictions of the $\Gamma(8)$ model, the corresponding KRR predictions given as translucent red square markers, lie further away from the identity mapping. 

Similar observations can be made for the case of $\Gamma(32)$ on the right-hand side pane of Figure \ref{fig_sortCM_questdb}. The best 10 predictions of the $\Gamma$-curve model are close to the identity line while the predictions of the KRR model are more loosely scattered. However, in this case, the poorest predicted molecules for single fidelity KRR do have some points close to the identity mapping line. Certainly, the overall scatter plot for $\Gamma(32)$ MFML looks closer to the identity mapping line as compared to the $\Gamma$(8) MFML model. Based on these results, one can hold on to what was stated before we started this assessment, the transferability of ML models remains a challenge, even for multifidelity approaches. Certainly the $\Gamma(8)$ MFML model is more efficient, and more accurate as has been sufficiently established in preceding sections. The key takeaway from this discussion on transferability is not the effectiveness of one model over the other but rather the fact that transferability is a difficult task for both single fidelity and multifidelity ML models.

\begin{figure}[htb!]
    \centering
    \includegraphics[width=\linewidth]{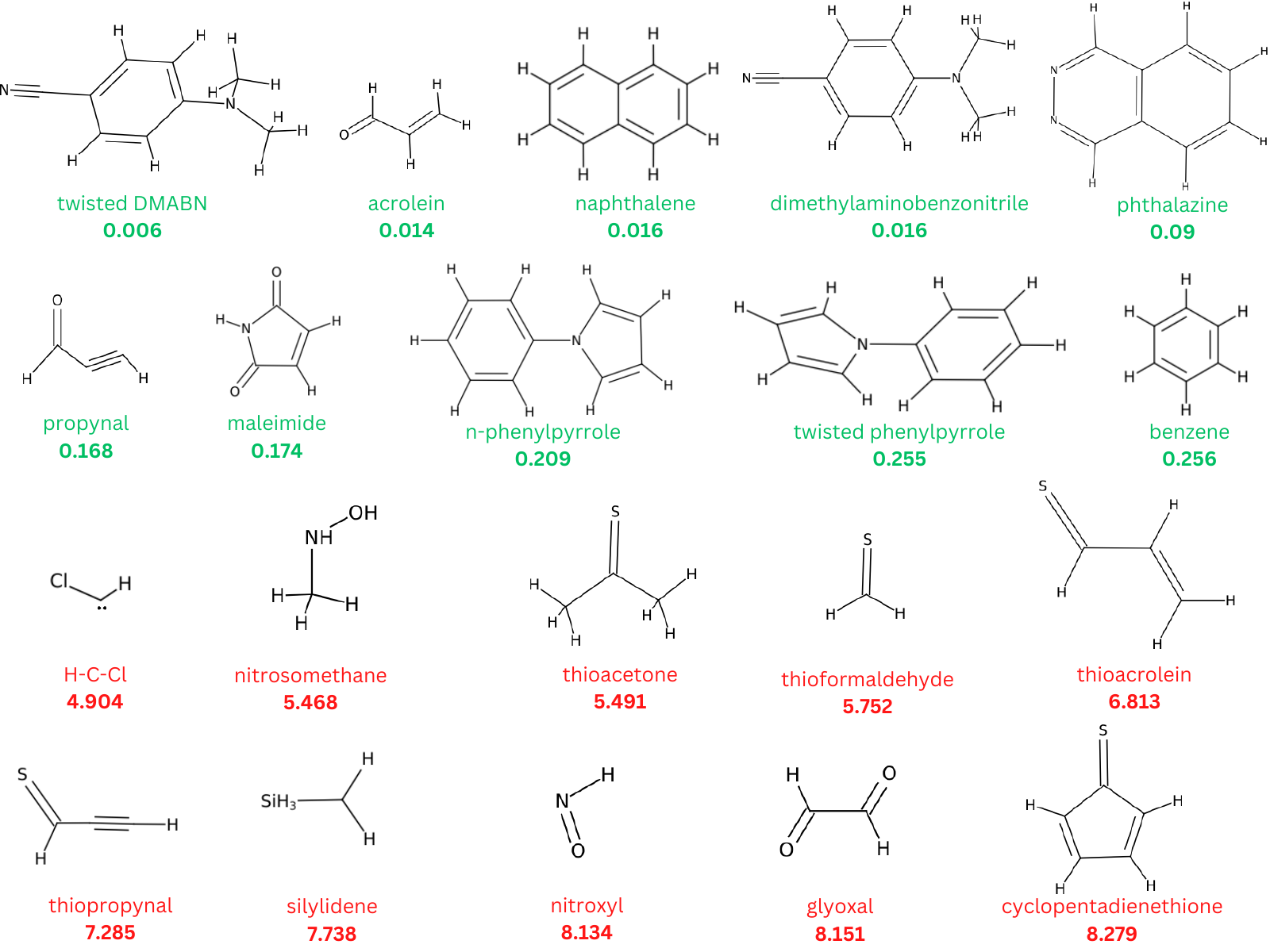}
    \caption{Best 10 (green) and worst 10 (red) predictions of the $\Gamma(32)$-MFML model over the QUESTDB dataset. The numbers under the name of the molecules indicate the relative error.}
    \label{fig_questdb_gamma32}
\end{figure}

To complete this discussion on transferability one can take a closer look into the errors of the $\Gamma(32)$ MFML model. Figure \ref{fig_questdb_gamma32} presents the molecules which comprise the 10 best predictions and 10 worst predictions along with the RE. It is unsurprising to notice that the DMABN and acrolein geometries from the QUESTDB database have well predicted energies since the QeMFi dataset, and by extension the training dataset, contains geometries of these molecules. As argued previously, this is due to the fact that the ML model trained on specific geometries does well in predicting the energies of similar geometries from `unseen' datasets. In addition, several other aromatic compounds such as naphthalene and phthalazine show low REs.
Consider the molecules that are not well predicted, the bottom two rows of Figure \ref{fig_questdb_gamma32}. 
One observes radicals and molecules with elements such as sulfur and silicon which are not present in the QeMFi dataset. Once again, it becomes clear how important the initial training dataset is to the ability of a ML model in predicting QC properties.

Concluding this digression on testing the $\Gamma$-curve for transferability, some final remarks are made here. As has become evident, the task of transferability of a ML model is challenging and demands investigation in its own right \cite{Zhang_2020_NN_tensorial_transfer}. The analysis of this short subsection is not indicative of an issue in the $\Gamma$-curve approach itself but rather ties into the larger picture of training general purpose ML models for QC \cite{krug2024a_foundational_model, batatia2024foundationmodelatomisticmaterials}. The reader is reminded that the work presented in this manuscript is intended to perform a training data hierarchy assessment for multifidelity ML methods in QC.

\section{Conclusions and Outlook}
This work discusses the concept of scaling for the number of training data across different fidelities for MFML and o-MFML in the prediction of excitation energies of the QeMFi dataset. Constant scaling factors, $\gamma$, were studied along with QC-calculation time-cost informed scaling factors, $\theta_{f-1}^f$ and $\theta_f^F$. It is seen in the results that the use of constant scaling factors,$\gamma$, is effective with a higher value of $\gamma$ resulting in lower model errors for reasonable time-cost of generating training data.
A new error metric, the error contour of MFML, was introduced and results discussed for the prediction of first vertical excitation energies of the QeMFi dataset. 
Such an analysis revealed that the data requirements for MFML-like methods is not as trivial as has been previously employed. In fact, one can achieve similar model accuracies with much less costly training samples if one increases the number of training samples at the lower end of the multifidelity data structure.
The error contours depicted that one could potentially use as little as 2 training sample at the target fidelities and achieve exceptional model accuracy if the subsequent fidelities used a larger number of training samples in comparison to MFML models built with some $\gamma$.
This was systematically studied with the newly introduced $\Gamma$-curve for a fixed number of training samples at the target fidelity and an increasing value of $\gamma$. The models built in this fashion were shown to be time-cost efficient over conventional MFML approach. A brief digression was made to thoroughly analyze transferability of the $\Gamma$-MFML model on completely unseen data from the QUESTDB database concluding that transferability is challenging for both single fidelity and multifidelity ML models.

These results provide a window into the inner mechanisms of MFML-like methods allowing for a better understanding of how they can be employed for accurate predictions of excitation energies with low cost of training data generation. 
The development of the $\Gamma$-curve approach in this work in its current form is only benchmarked for a specific DFT functional and this could be a potential limitation of this work. A possible extension of this work could be the study of the $\Gamma$-curve approach for a wider range of DFT functionals. At the moment this is inhibited by the lack of compute cost times in most large-scale multifidelity datasets. Another interesting area of research can be the use of approaches developed in this work to assess the efficiency of multifidelity approaches for fidelity structures built on CC level of theory and would be of particular interest since CC is considered the gold standard in QC.
Future methodological development over this could include algorithms that optimally select the number of training samples to be used at each fidelity. Further work could involve extending this to other time-based scaling factors which also take into account the error reduction that each additional training sample contributes to the overall multifidelity model. 
This would further improve the scope of application for the multifidelity methods discussed herein.

\subsubsection*{Data availability}
The data used in this study was taken from the QeMFi dataset which can be found in \href{https://zenodo.org/records/13925688}{this ZENODO data repository}. 

\subsubsection*{Code availability}
The programming scripts used for this study can be openly accessed at \href{https://github.com/SM4DA/MFML_DataHierarchy} {this GitHub repository}.

\subsubsection*{Acknowledgments}
The authors acknowledge support by the DFG through the project ZA 1175/3-1 as well as through the DFG Priority Program SPP 2363 on “Utilization and Development of Machine Learning for Molecular Applications – Molecular Machine Learning” through the project ZA 1175/4-1. The authors would also like to acknowledge the support of the `Interdisciplinary Center for Machine Learning and Data Analytics (IZMD)' at the University of Wuppertal.

\subsubsection*{Declarations}
The authors declare that there is no conflict of interest or any competing interests.

\bibliography{main}

\end{document}